\documentclass[aps,pra,epsfig]{revtex4}

\usepackage[dvips]{graphics}
\usepackage{graphicx}
\usepackage{amsfonts}
\usepackage{amssymb}
\usepackage{amsmath}

\begin{document}

\title{Ultracold Bosons with $3$-Body Attractive Interactions in an Optical
Lattice}
\author{E. Fersino$^{1,2}$, B. A. Malomed$^{3}$, 
G. Mussardo$^{1,2,4}$, and A. Trombettoni$^{1,2}$}
\affiliation{
$^1$ International School for Advanced Studies, Via Beirut 2/4, Trieste 34014,
Italy\\
$^2$ Istituto Nazionale di Fisica Nucleare, Sezione di Trieste, Italy\\
$^3$ Department of Physical Electronics, School of Electrical Engineering,
Faculty of Engineering, Tel Aviv University, Tel Aviv 69978, Israel\\
$^4$ 
Abdus Salam International Centre of Theoretical Physics, Strada Costiera 11,
Trieste 34014, Italy}


\begin{abstract}
We study the effect of an optical lattice (OL) on 
the ground-state properties of one-dimensional ultracold bosons 
with three-body attraction
and two-body repulsion, which are described by a
cubic-quintic Gross-Pitaevskii equation with a periodic potential. Without
the OL and with a vanishing two-body interaction term, 
soliton solutions of the Townes type are
possible only at a critical value of the three-body interaction strength, 
at which an infinite degeneracy of the ground-state occurs; 
a repulsive two-body interaction makes 
such localized solutions unstable. 
We show that the OL opens a stability window around 
the critical point when the strength of the periodic potential is above
a critical threshold.
We also consider the effect of an external parabolic trap, studying 
how the stability of the solitons depends on matching 
between minima of the periodic potential and the minimum 
of the parabolic trap.
\end{abstract}
%

\maketitle

\section{Introduction}

\label{intro}

In the current studies of ultracold quantum gases, a great deal of interest
has been drawn to the study of Bose-Einstein condensates (BECs) loaded into
optical lattices (OLs), i.e., spatially periodic potentials induced by the
interference between counterpropagating laser beams \cite%
{book1,book2,morsch06}. Besides playing a crucial role in effectively tuning
the interaction strength in the condensate, i.e., the ratio between the
kinetic and interaction energies \cite{jaksch98}, OLs offer an extremely
useful tool for studies of the transition between the superfluid and
Mott-insulator states \cite{hansch02}, and for the investigation of effects
in the matter-wave dynamics due to the interplay between nonlinearity and
the quasi-discreteness, which is induced by a deep lattice potential
\cite{morsch06}.

The mean-field dynamics of the BEC loaded into the OL is described by the
cubic Gross-Pitaevskii equation (GPE) with the periodic potential \cite%
{book1,book2,morsch06}. The respective Bogoliubov's excitation spectrum
features a band structure, similar to electronic Bloch bands in solid state.
If the OL potential is deep enough, the lowest-band dynamics may be
approximated by the discrete nonlinear Schr\"{o}dinger (NLS) equation \cite%
{trombettoni01}. Using this correspondence, the BEC\ dynamics was studied in
the framework of the nonlinear-lattice theory, see works \cite%
{trombettoni01,abdullaev01,alfimov02} and short review \cite{Mason}. The
presence of the OL gives raise to energetic and dynamical
instabilities, which have been predicted theoretically \cite%
{wu01,smerzi02,konotop02,wu03,menotti03,taylor03,kramer03,kramer05} and
studied experimentally \cite{cataliotti03,inguscio04}.

An important application of the OLs is their use for the creation and
stabilization of matter-wave solitons. In particular, the periodic potential
gives rise to localized gap solitons in the case of \emph{repulsive}
two-body interactions, as was predicted theoretically \cite{GSprediction}
and demonstrated experimentally \cite{eiermann04} (with the attractive
interactions, bright matter-wave solitons were created and observed in
condensates of $^{7}$Li \cite{strecker02,khaykovich02} and $^{85}$Rb \cite%
{Weiman} atoms). More generally, the use of time- and space-modulated fields
acting on atoms is a powerful tool for the control of soliton properties
\cite{malomed06}; for instance, while the GPE without external potentials
admits stable soliton solutions only in the 1D geometry \cite%
{sulem99,ablowitz04}, OL potentials can stabilize solitons in any higher
dimension \cite{malomed03,Yang,BBB,Barcelona}. Unlike 1D solitons, a necessary
existence condition for their multidimensional counterparts, stabilized by
means of OLs, is that the soliton's norm must exceed a certain threshold
value.

Another very useful tool frequently used in experiments with ultracold
atomic gases is the control of the strength and sign of two-body
interactions by means of an external magnetic field near the Feshbach
resonance \cite{book1,book2}. Further, recent works proposed to exploit the
possibility to control the strength of \emph{three-body} interactions
between atoms, independently from the control of the two-body collisions
\cite{paredes07,buchler07}. One motivation for such studies is related to
the possibility of creating new exotic strongly correlated phases in
ultracold gases. Indeed, quantum phases, such as topological ones or spin
liquids, turn out to be ground-states of the Hamiltonian including three- or
multi-body-interaction terms, an example being fractional quantum-Hall
states described by Pfaffian wave functions \cite{moore91}. In a recent work
\cite{fersino08}, a 1D Bose gas with $\mathcal{N}$-body attractive
interactions was studied in the mean-field approximation, with the objective
to create highly degenerate ground-states of Hamiltonians including
many-body terms. For the three-body interactions ($\mathcal{N}=3$), the
system is described by a \emph{quintic} GPE, i.e., the respective term in
the energy density is proportional to $|\psi |^{6}$, where $\psi $ is the
single-atom mean-field wave function (in the general case, a similar term is
proportional to $|\psi |^{2\mathcal{N}}$).

Soliton solutions can be found for each $\mathcal{N}$, but they represent
the stable ground-state, with negative energy (which is defined as per Eqs. (%
\ref{E-functional}) and (\ref{h0}), see below), only for $\mathcal{N}=2$,
being unstable excited states with positive energy at $\mathcal{N}\geq 4$.
For $\mathcal{N}=3$, soliton solutions are 1D counterparts of the well-known
Townes solitons \cite{abdullaev05}, which play the role of the separatrix
between collapsing and decaying localized states. 
The Townes-like solitons with fixed norm (which is $1$, in
the notation adopted below) exist only at a single critical value of the
interaction strength, at which they feature the infinite degeneracy \cite%
{fersino08}: \emph{all} the solitonic wave functions, $\psi (x)=\mathrm{const%
}\cdot \left[ \sigma \cosh {\left( x/\sigma \right) }\right] ^{-1/2}$, with
arbitrary width $\sigma $ (see Eqs. (\ref{critical-value}) and (\ref%
{quintic-1}) below), have \emph{zero energy} but different values of the
chemical potential, $\mu \sim \sigma ^{-1}$. A relevant issue is how this
infinite degeneracy is lifted by an external potential, especially by a
periodic one corresponding to the OL \cite{malomed03,abdullaev05}.

When the two-body interaction is present, the mean-field equation
is the GPE with the cubic-quintic (CQ) nonlinearity
\cite{abdullaev05,khaykovich06}. As said above, it has been shown
\cite{buchler07} that it is possible to tune the strength of the
two-body interactions independently from the three-body ones. In
addition to that, in the framework of the effective GPE for the
BEC loaded into a nearly 1D (``cigar-shaped") trap with tight
transverse confinement, an effective attractive quintic term
appears, in the absence of any three-atom interactions, as a
manifestation of the residual deviation from the
one-dimensionality \cite{nearly1D,khaykovich06}. In any case, if
the two-body interaction is repulsive while its three-body
counterpart is attractive, soliton solutions to the CQ GPE can be
found in an exact analytical form (in the free space), but they
feature an unstable eigenvalue in the Bogoliubov - de Gennes
spectrum of small perturbations around them \cite{khaykovich06},
while the instability of the Townes-like solitons in the quintic
equation is subexponential, being accounted for by a zero
eigenvalue. 

The issue we address in this paper is the possibility
to stabilize such solitons by means of the OL potential.
Previously, the stabilization of originally unstable solitons by
means of the OL was considered, in the 2D
\cite{malomed03,Yang,BBB} and 1D \cite{abdullaev05} settings
alike, only for localized states of the Townes type (recently, the
stabilization of 2D solitons against the \textit{supercritical
collapse}  by the OL was also demonstrated in the CQ model in 2D,
with both cubic and quintic terms being attractive \cite{Radik2}).
It was found that the OL with
any value of its strength (i.e., with zero threshold) opens a \textit{%
stability window} around the critical point corresponding to the Townes
solitons. In this work, we demonstrate that the OL opens a stability window
for solitons in the CQ model (with the repulsive cubic and attractive
quintic terms) too, but only if the lattice strength exceeds a \emph{finite}
threshold value.

Apart from the context of BEC, where the nonlinearity degree is related to
the number of atoms simultaneously involved in the contact interaction, NLS
equations with the power-law and CQ nonlinearities are also known as
spatial-domain models of the light propagation in self-focusing media \cite%
{kivshar03} (for a brief overview of optical models based on the CQ-NLS
equation, including references to experimental realizations, see recent
works \cite{Radik2,Radik1}). In the case of the cubic nonlinearity (the Kerr
medium), effects of imprinted lattices on the transmission of light beams
have been investigated both in local \cite{malomed99,malomed03} and nonlocal
\cite{xu05,wang08} models.

The paper is structured as follows. In Section II, we introduce the CQ GPE
corresponding to the mean-field description of the 1D Bose gas with two-body
repulsive and three-body attractive interactions. Properties of the
(unstable) soliton solutions to this equation are also recapitulated in
Section II. In Section III, we use the variational approximation (VA) (see
Ref. \cite{malomed01} for a review) to discuss effects of the OL on the
solitons. We introduce an appropriate ansatz and compute the corresponding
energy. The limit of the vanishing two-body interaction is considered too
and compared to previous results \cite{abdullaev05}. In Section IV, the
stability region for the soliton solution in the presence of the repulsive
two-body interaction and OL is determined and compared with numerical
findings. The effect of an additional harmonic-trap potential is studied in
Section V, showing that the stability region depends on the matching between
minima of the periodic potential and the location of the minimum of the
harmonic trap. In Section VI we present our conclusions.

\section{The model}

\label{model}

The quantum many-body Hamiltonian for the 1D Bose gas with $\mathcal{N}$%
-body contact attractive interactions is
\begin{equation}
\hat{H}=\int_{-\infty }^{+\infty }dx\left\{ \hat{\Psi}^{\dag }(x)\hat{h}_{0}%
\hat{\Psi}(x)-\frac{c}{\mathcal{N}!}\left[ \hat{\Psi}^{\dag }(x)\right] ^{%
\mathcal{N}}\left[ \hat{\Psi}(x)\right] ^{\mathcal{N}}\right\} ,
\label{Ham-contact}
\end{equation}%
where $\hat{\Psi}(x)$ is the bosonic-field operator, $c>0$ is the
nonlinearity strength and
\begin{equation}
\hat{h}_{0}=-\frac{\hbar ^{2}}{2m}\,\frac{\partial ^{2}}{\partial x^{2}}+V_{%
\mathrm{ext}}(x)  \label{h0}
\end{equation}%
is the single-particle Hamiltonian, $V_{\mathrm{ext}}(x)$ being the external
potential. The case of $\mathcal{N}=2$ in the homogeneous limit ($V_{\mathrm{%
ext}}=0$) corresponds to the integrable Lieb-Liniger model \cite{lieb63}.
For attractive interactions ($c>0$), its analytical solution was obtained by
means of the Bethe ansatz \cite{mcguire64} and for a large number of
particles, $N_{\mathrm{tot}}$, the energy of the exact ground-state solution
coincides with that obtained in the mean-field approximation \cite%
{calogero75}. In the attractive Lieb-Liniger model, a finite ground-state
energy per particle is provided by fixing product $cN_{\mathrm{tot}}$ to a
constant value \cite{mcguire64,calogero75}, while for $\mathcal{N}>2$ one
has to set $c\left( N_{\mathrm{tot}}\right) ^{\mathcal{N}-1}=\mathrm{const}$
\cite{fersino08}.

In the Heisenberg representation, the equation of motion for field $\hat{\Psi%
}(x,t)$ is
\begin{equation}
i\hbar \frac{\partial \hat{\Psi}}{\partial t}=\left[ \hat{\Psi},\hat{H}%
\right] =\hat{h}_{0}\hat{\Psi}-c\left( \hat{\Psi}^{\dag }\right) ^{\mathcal{N%
}-1}\left( \hat{\Psi}\right) ^{\mathcal{N}-1}\hat{\Psi}.  \label{dyn}
\end{equation}%
The mean-field approximation reduces Eq. (\ref{dyn}) to the corresponding
GPE with the power-law nonlinearity,
\begin{equation}
i\hbar \frac{\partial \psi (x,t)}{\partial t}=\left( \hat{h}_{0}-c|\psi
(x,t)|^{\alpha }\right) \psi (x,t),  \label{GNLS}
\end{equation}%
where the macroscopic wave function $\psi (x,t)$ is normalized to the total
number of atoms, $N_{\mathrm{tot}}$, and the nonlinearity degree is related
to the order of the multi-body interactions, $\mathcal{N}$:
\begin{equation}
\alpha =2\left( \mathcal{N}-1\right) .  \label{alpha}
\end{equation}%
Thus, the usual two-body interaction ($\mathcal{N}=2$) corresponds to $%
\alpha =2$, and the three-body interaction ($\mathcal{N}=3$) to $\alpha =4$.
Equation (\ref{GNLS}) conserves the energy,
\begin{equation}
E=\int dx\psi ^{\ast }(x)\left[ \hat{h}_{0}-\frac{2c}{\alpha +2}|\psi
(x)|^{\alpha +2}\right] \psi (x),  \label{E-functional}
\end{equation}%
which is the classical counterpart of quantum Hamiltonian (\ref{Ham-contact}%
).

In Eq. (\ref{h0}), $V_{\mathrm{ext}}(x)$ is the external trapping potential,
which typically includes a superposition of an harmonic magnetic 
trap and periodic OL potential, $V_{\mathrm{ext}%
}(x)=V_{\mathrm{HO}}(x)+V_{\mathrm{OL}}(x)$, where the harmonic confining
term is $V_{\mathrm{HO}}=
m\omega ^{2}x^{2}/2$. We take the periodic potential as $V_{%
\mathrm{OL}}=\epsilon \sin ^{2}{(qx+\delta )}$, where $\epsilon $ is
proportional to the power of the laser beams which build the OL, and $q=2\pi
/\lambda $, with $\lambda =\lambda _{\mathrm{laser}}\sin {\left( \theta
/2\right) }${;} here, $\lambda _{\mathrm{laser}}$ is the wavelength of the
beams, and $\theta $ the angle between them (the period of the lattice is $%
\lambda /2$). Parameter $\delta $ measures a mismatch between the minimum of
the parabolic potential (at $x=0$) and the closest local minimum of the
lattice potential: when $\delta =0$ ($\delta =\pi /2$) a minimum (maximum)
of $V_{\mathrm{OL}}$ coincides with the minimum of $V_{\mathrm{HO}}$. In
fact, except for Section V, we consider the situation without the parabolic
trap (i.e., $\omega =0$), therefore we set $\delta =0$ in this case.

The time-independent power-law GPE corresponding to Eq. (\ref{GNLS}) is
(from now on, we use normalized units, with $\hbar =m=1$ and $N_{\mathrm{tot}%
}=1$)
\begin{equation}
\left[ -\frac{1}{2}\frac{d^{2}}{dx^{2}}-c|\psi (x)|^{\alpha }+V_{\mathrm{ext}%
}(x)\right] \psi (x)=\mu \psi (x),  \label{GNLS-mu}
\end{equation}%
where $\mu $ is the chemical potential,
\begin{equation}
V_{\mathrm{ext}}(x)=\epsilon \sin ^{2}{(qx)},  \label{OL}
\end{equation}
and the norm of the wave function is $1$. In the free-space cubic model ($%
\alpha =2$ and $\epsilon =0$), Eq. (\ref{GNLS-mu}) is the integrable NLS
equation, whose multi-soliton solutions can be obtained by means of the
inverse scattering method \cite{ablowitz04}. The commonly known
single-soliton NLS solution is
\begin{equation}
\psi (x)=A~\mathrm{sech}\left( kx\right) ,  \label{cubic-1}
\end{equation}%
where $k^{2}\equiv 2|\mu |$ and $A$ is a real amplitude, the respective
value of the chemical potential being $\mu =-cA^{2}/2$. For a general value
of $\alpha >0$, the integrability is lost even in the absence of the
external potential \cite{sulem99}; nevertheless, the respective
single-soliton solutions can be found in an explicit form \cite%
{ablowitz81,polyanin04}.

For the attractive three-body interactions ($\alpha =4$), Eq. (\ref{GNLS})
is the self-focusing quintic GPE, whose stationary version is
\begin{equation}
\left[ -\frac{1}{2}\frac{d^{2}}{dx^{2}}-c|\psi (x)|^{4}+V_{\mathrm{ext}}(x)%
\right] \psi (x)=\mu \psi (x).  \label{quintic-mu}
\end{equation}%
For $V_{\mathrm{ext}}(x)=0$, 
if one fixes coefficient $c$ in front of the interaction term, the
Townes-like solitons exist for a particular value of the norm of the wave 
function \cite{gaididei99,abdullaev05}. On the other hand, fixing the 
normalization of the wave function (recall that the norm is $1$ in our 
units) amounts, for $\alpha \neq 4$, to fixing a relation among the chemical 
potential and the interaction strength \cite{fersino08}, so that for each $c$ 
it is possible to obtain a single soliton solution (although, as mentioned 
above, these solutions provide the ground-state in the infinite system only
for $\alpha <4$, i.e., for $\mathcal{N}<3$). However, for $\alpha =4$ (i.e.,
$\mathcal{N}=3$) chemical potential $\mu $ remains indefinite, assuming
arbitrary negative values, while the soliton solution of the form
\begin{equation}
\psi (x)=\frac{\left( 3k^{2}/8c\right) ^{1/4}}{\sqrt{\cosh (kx)}}%
,\,~k^{2}=-8\mu   \label{quintic-1}
\end{equation}%
satisfies the unitary normalization condition at a single (critical) value
of the interaction strength \cite{gaididei99,fersino08},
\begin{equation}
c=c^{\ast }\equiv \frac{3\pi ^2}{8}.  \label{critical-value}
\end{equation}%
At $c=c^{\ast }$, all solutions (\ref{quintic-1}) share a common value of
the energy, which is simply $E=0$ \cite{abdullaev05,fersino08}, as follows
from Eqs. (\ref{E-functional}) and (\ref{critical-value}).

If the two-body interaction is added to the three-body attraction, the
mean-field equation is the GPE with the CQ nonlinearity,
\begin{equation}
\left[ -\frac{1}{2}\frac{d^{2}}{dx^{2}}+g|\psi (x)|^{2}-c|\psi (x)|^{4}+V_{%
\mathrm{ext}}(x)\right] \psi (x)=\mu \psi (x).  \label{cubic-quintic-mu}
\end{equation}%
As said above, we chiefly focus on the case of the \emph{repulsive} two-body
interactions, i.e., $g\geq 0$. A family of exact soliton solutions to Eq. (%
\ref{cubic-quintic-mu}) with $V_{\mathrm{ext}}(x)=0$ can be obtained in the
exact form \cite{pushkarov79,cowan86,khaykovich06}, which, for $g\geq 0$, is
\begin{equation}
\psi ^{2}(x)=\frac{A^{2}}{(1+\xi A^{2})\cosh {(2\sqrt{2|\mu |}x)}-\xi A^{2}}%
~,  \label{cubic-quintic-1}
\end{equation}%
where $\xi \equiv g/\left( 4|\mu |\right) $, and the maximum value of the
density, at the soliton's center, is
\begin{equation}
A^{2}=\frac{3}{c}\left( \frac{g}{4}+\sqrt{\frac{g^{2}}{16}+\frac{c|\mu |}{3}}%
\right) .  \label{A-cubic-quintic}
\end{equation}%
A simple derivation of Eq. (\ref{cubic-quintic-1}) is presented in Appendix
A. Obviously, for $g=0$ solution (\ref{cubic-quintic-1}) reduces to
Townes-like soliton (\ref{quintic-1}).

Imposing the above-mentioned normalization,%
\begin{equation}
\int_{-\infty }^{+\infty }|\psi (x)|^{2}dx=1,  \label{1}
\end{equation}%
on solution (\ref{cubic-quintic-1}), one arrives at relation%
\begin{equation}
\sqrt{\frac{6}{c}}\tan ^{-1}\left( \sqrt{1+2\xi A^{2}}\right) =1,
\label{norm-cubic-quintic}
\end{equation}%
from where it follows that, for $g>0$, soliton solutions with $\mu <0$
satisfying normalization condition (\ref{1}) exist for $c>c^{\star }$.
However, these solutions are unstable \cite{khaykovich06} (in particular,
because they do not satisfy the Vakhitov-Kolokolov stability criterion \cite%
{kolokolov73}). In the following section we discuss how the OL can stabilize
such localized solutions.

\section{Variational approximation}

\label{variational}

Both for $\alpha =2$ and $4$ ($\mathcal{N}=2$ and $3$), and for the GPE with
the mixed CQ nonlinearity, the presence of the periodic potential makes it
necessary to resort to approximate methods for finding solitons. To this
end, we use the VA (variational approximation) \cite{baym96,malomed01}
based on the \emph{ansatz} which yields exact soliton solution (\ref%
{quintic-1}) of the quintic NLS equation in the absence of the external
potential:
\begin{equation}
\psi _{\mathrm{ans}}(x)=\frac{A}{\sqrt{\cosh (x/\sigma )}}.  \label{VAR}
\end{equation}%
Here, width $\sigma $ is the variational parameter to be determined by the
minimization of the energy, while amplitude $A$ will be found from
normalization condition (\ref{1}). We expect that ansatz (\ref{VAR}), which
does not explicitly include the modulation of the wave function induced by
the OL, may give a reasonable estimate of the soliton's energy for
sufficiently small values of OL strength $\epsilon $ in Eq. (\ref{OL}), cf.
the known result for the 2D equation with the cubic nonlinearity ($\alpha =2$%
) and OL potential \cite{malomed03,BBB}. In the case of the 3D GPE which
includes the cubic term and harmonic trap, this approach leads to an
estimate for the critical value of the number of atoms above which the
condensate collapses, that was found to be in a reasonable agreement with
results produced by the numerical solution of the GPE \cite%
{fetter95,ruprecht95}. In 1D, the VA based on the Gaussian ansatz also 
provides for quite an accurate approximation 
to exact soliton solution (\ref{cubic-1} \cite{Anderson}. 
Similar analyses carried out in the 1D model
including the cubic term and OL \cite{malomed99,malomed03,delocalization}
have demonstrated that (unlike the 2D and 3D cases) the 1D soliton trapped
in the OL potential does not have an existence threshold in terms of its
norm (number of atoms).

The energy to be minimized in the framework of the VA is obtained by
inserting ansatz (\ref{VAR}) in the GPE energy functional given by Eq. (\ref%
{E-functional}). The kinetic and quintic-interaction energy terms in the
functional both scale as $\sigma ^{-2}$; then, the energy per particle
computed from expression (\ref{E-functional}) is
\begin{equation}
E=\frac{\beta }{\sigma ^{2}}+\frac{g}{\pi ^{2}\sigma }+\frac{\epsilon }{2}%
\left[ 1-\mathrm{sech}(\pi q\sigma )\right] ,  \label{E3}
\end{equation}%
\begin{equation}
\beta \equiv \frac{1}{16}-\frac{c}{6\pi ^{2}}=\frac{c^{\ast }-c}{6\pi ^{2}},
\label{beta}
\end{equation}%
where $c^{\ast }$ is defined in Eq. (\ref{critical-value}).

For $\epsilon =0$ (without the OL), the scenario discussed in the previous
section for the uniform CQ GPE with the attractive three-body and repulsive
two-body interactions is recovered, as energy (\ref{E3}) reduces in that
case to
\begin{equation}
E=\frac{\beta }{\sigma ^{2}}+\frac{g}{\pi ^{2}\sigma }.  \label{E3-homog}
\end{equation}%
For $g=0$, the energy is positive when $c<c^{\ast }$ (i.e., $\beta >0$, see
Eq. (\ref{beta})) and vanishes at $\sigma \rightarrow \infty $; for $%
c=c^{\ast }$ (i.e., $\beta =0$) one obtains $E=0$, in agreement with the
above-mentioned exact result showing the infinite degeneracy of soliton
family (\ref{quintic-1}), while for $c>c^{\ast }$ the energy is negative and
diverges (to $-\infty $) at $\sigma \rightarrow 0$, signaling, in terms of
the VA, the onset of the collapse. With $g>0$, expression (\ref{E3-homog})
does not give rise to any minimum of the energy, which agrees with the known
fact of the instability of all the solitons in this case \cite{khaykovich06}.

A detailed study of minima of variational energy (\ref{E3}) is presented in
Appendix B. In the following subsection, we consider the case of the
self-focusing quintic GPE in the presence of the OL ($\epsilon >0,~g=0$),
while the discussion of the general case ($\epsilon >0,~g>0$) is given in
Section IV.

\subsection{Self-focusing quintic GPE with the optical-lattice potential}

\label{pure_quintic}

Here we address the stability of localized variational mode (\ref{VAR}), for
different values the OL parameters, strength $\epsilon $ and wavenumber $q$,
keeping $g=0$. The results of the analysis of minima of the variational
energy (\ref{E3}), presented in Appendix B, can be summarized as follows
(see also Fig. \ref{fig1}): for $c\geq c^{\ast }$, the infinitely deep
minimum of the energy is obtained at $\sigma \rightarrow 0$, which
corresponds to the collapse, as shown in Fig. \ref{fig1}(a). For $c<c^{\ast }
$, the collapse may be avoided, and three possibilities arise: there exists
another special value, $c^{\prime }<c^{\ast }$, such that for every $c$
between $c^{\prime }$ and $c^{\ast }$ the energy has a minimum at $\sigma
=\sigma _{1}$ and a maximum at $\sigma =\sigma _{2}$, while for $c<c^{\prime
}$ the energy does not have a minimum at any finite value of $\sigma $, see
Fig. \ref{fig1}(d). Actually, two different situations should be
distinguished for $c^{\prime }<c<c^{\ast }$: there exists a specific value
(refer to Appendix B),
\begin{equation}
c^{\ast \ast }=c^{\ast }-\frac{3\epsilon }{2q^{2}}T_{c},\mathrm{\ where~}%
T_{c}\approx 2.13,  \label{c_2_star}
\end{equation}%
(with $c^{\ast \ast }>c^{\prime }$) such that, for $c^{\ast \ast }<c<c^{\ast
}$, the energy has a \emph{global} minimum at $\sigma =\sigma _{1}$ (which,
thus, represents the \emph{ground-state} of the boson gas in this
situation), while, for $c^{\prime }<c<c^{\ast \ast }$, the energy minimum at
$\sigma =\sigma _{1}$ is a \emph{local} one. In other words, taking into
regard the fact that, as shown by Eq. (\ref{E3}), the energy-per-particle
approaches value $\epsilon /2$ at large $\sigma $, we conclude that, for $%
c^{\ast \ast }<c<c^{\ast }$ ($c^{\prime }<c<c^{\ast \ast }$ ), the energy
satisfies inequality $E(\sigma _{1})<\epsilon /2$ ($E(\sigma _{1})>\epsilon
/2$), as showed in Figs. \ref{fig1}(b,c).

\begin{figure}[tbp]
\begin{center}
\includegraphics[width=6.cm,height=6.cm,angle=270,clip]{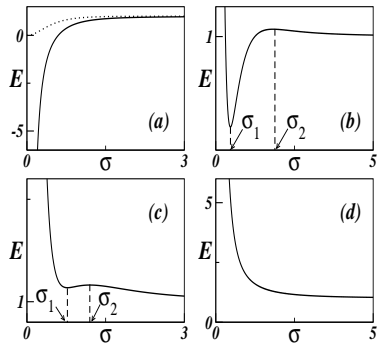}
\end{center}
\caption{Variational energy $E$ obtained in the framework of the quintic GPE
versus $\protect\sigma $ (in units of $\protect\epsilon /2$) for $c\geq
c^{\ast }$ (a); $c^{\ast \ast }<c<c^{\ast }$ (b); $c^{\prime }<c<c^{\ast
\ast }$ (c); $c<c^{\prime }$ (d). In (a) the solid (dotted) line is the
energy for $c>c^{\ast }$ ($c=c^{\ast }$); in (b)-(c), points of the energy
minimum and maximum, $\protect\sigma _{1}$ and $\protect\sigma _{2}$, are
indicated.}
\label{fig1}
\end{figure}

From the above analysis, we infer that for $c<c^{\ast \ast}$
the ground-state is a delocalized one (although the metastable state,
corresponding to the above-mentioned local energy minimum, exists for
$c^{\prime }<c<c^{\ast \ast }$), for $c^{\ast \ast }<c<c^{\ast }$ the
ground-state is represented by a finite-size soliton configuration (in
agreement with Ref. \cite{abdullaev05}) and for $c>c^{\ast }$ it is
collapsing. Equation (\ref{c_2_star}) 
shows that the width of the stability region depends on ratio $%
\epsilon /q^{2}$: keeping fixed all other parameters, the decrease of the
lattice spacing (i.e., the increase of $q$) leads to a reduction of the
stability region. Equation (\ref{c_2_star}) also shows that for $\epsilon
/q^{2}=2c^{\ast }/3T_{c}\approx 1.16$ the VA formally predicts $c^{\ast \ast
}=0$: however, for $c=0$, the ground-state is delocalized and the
variational ansatz (\ref{VAR}) cannot be used, as it does not take into
account the modulation induced by the deep OL potential.

\begin{figure}[t]
\begin{center}
\includegraphics[width=6.cm,height=6.cm,angle=270,clip]{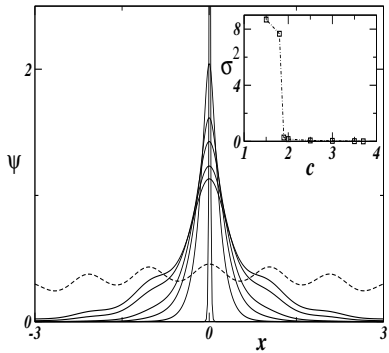}
\end{center}
\caption{The numerically found ground-state of the quintic GPE with the
periodic OL potential, for several values of nonlinearity coefficient $c$.
Solid lines, starting from the narrowest configuration, refer to $%
c=3.7,3.5,3,2.5,2,1.9$ (recall that $c^{\ast }=3\protect\pi ^{2}/8\simeq
3.701$), and the dashed line refers to $c=1.8$. 
Parameters are $\protect\epsilon =6$, $q=3$ and $L=10$. Inset: squared width $\protect\sigma^2$
of the ground-state as a function of $c$ (the dot-dashed line is a guide to
the eye). Critical value $c^{\ast \ast }$ obtained from the numerical
analysis is $c^{\ast \ast }=1.87(3)$, which should be compared with 
the corresponding value (\protect\ref{c_2_star}) 
predicted by the variational approximation, $c^{\ast
\ast }\simeq 1.57$.}
\label{deloc-1}
\end{figure}

\begin{figure}[t]
\begin{center}
\includegraphics[width=6.cm,height=6.cm,angle=270,clip]{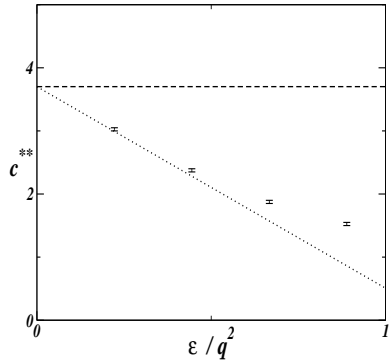}
\end{center}
\caption{The dotted line: the variational estimate for $c^{\ast \ast }$ as a
function of $\protect\epsilon /q^{2}$, according Eq. (\protect\ref{c_2_star}%
) (for the quintic GPE), the dashed line corresponding to $c^{\ast }=3%
\protect\pi ^{2}/8$. Discrete symbols represent results obtained from the
numerical solution of the quintic GPE. They designate the transition form
the localized ground-state to the extended one (parameters are the same as
in Fig. (\protect\ref{deloc-1}). According to the variational approximation,
the ground-state is delocalized ($\protect\sigma \rightarrow \infty $) below
the dotted line, and it collapses ($\protect\sigma \rightarrow 0$) for $c$
above the dashed line.}
\label{comparison_quintic}
\end{figure}

In Fig. \ref{deloc-1}, we plot the numerically found ground-state of the
quintic GPE in a 1D box ($-L<x<L$). It is seen that, with the increase of $%
c^{\ast }-c$, the configuration becomes broader, until a critical value is
reached, as discussed in \cite{abdullaev05}. 
In the inset of Fig. \ref{deloc-1} we plot the squared 
width $\sigma^2=\int_{-\infty}^\infty dx \, x^2 \mid \psi(x)\mid^2$ 
of the numerically found ground-state $\psi$ versus $c$, 
which makes the delocalization transition evident: for $c<c^{\ast \ast}$ 
the width $\sigma$ is $ \propto L$, while around $c \approx c^{\ast \ast}$ 
the width suddenly decreases. 
Variational estimate (\ref{c_2_star}) for the critical value $c^{\ast \ast}$, as predicted by the
VA (see Eq. (\ref{c_2_star})), is displayed in Fig. \ref{comparison_quintic}, 
together with numerical results. One observes a reasonable agreement
between them, especially for small $\epsilon $, which is due both to the use
of the more adequate ansatz (\ref{VAR}), rather than a Gaussian, and also
because $T_{c}$ is found as the value at which the global (rather than
local) minimum disappears.

\section{The stability region for the condensate with competing two- and
three-body interactions}

\label{three_body}

The most interesting situation occurs when the two-body repulsive
interaction ($g>0$) competes with the attractive three-body collisions 
($c>0$). As said above, all solitons in the free space ($\epsilon =0$) are
strongly unstable in this situation \cite{khaykovich06}, and the possibility
of their stabilization by the OL was not studied before. The analysis of
variational energy (\ref{E3}), presented in Appendix B, yields the following
results for this case. For $c>c^{\ast }$, the energy does not have a minimum
at finite $\sigma $, hence the OL cannot stabilize the solitons in this
case. If $c=c^{\ast }$, the energy has a global minimum at a finite value of
$\sigma $, when
\begin{equation}
G\equiv \frac{2gq}{\pi \epsilon }<G_{c}\approx 0.663.  \label{G-cond-text}
\end{equation}%
For $c<c^{\ast }$, the energy features a global minimum at finite $\sigma $
for $c^{\ast \ast }(G)<c<c^{\ast }$, where the modified critical value is
\begin{equation}
c^{\ast \ast }(G)\equiv c^{\ast }-\frac{3\epsilon }{2q^{2}}T_{c}(G),
\label{c_2_star_G}
\end{equation}%
cf. definition (\ref{c_2_star}) for $G=0$. The value $T_{c}$ depends upon $G$%
, vanishing for $G$ larger than the critical value $G_{c}$. This means that,
to balance the destabilizing effect of the repulsive two-body interactions,
the strength of the periodic potential, $\epsilon $, must \emph{exceed} its
own critical value,
\begin{equation}
\epsilon _{\mathrm{crit}}=\frac{2qg}{\pi G_{c}}.  \label{epsilon_critical}
\end{equation}%
Otherwise, Eq. (\ref{c_2_star_G}) yields $c^{\ast \ast }=c^{\ast }$, i.e.,
the OL cannot stabilize the solitons.

\begin{figure}[t]
\begin{center}
\includegraphics[width=6.cm,height=6.cm,angle=270,clip]{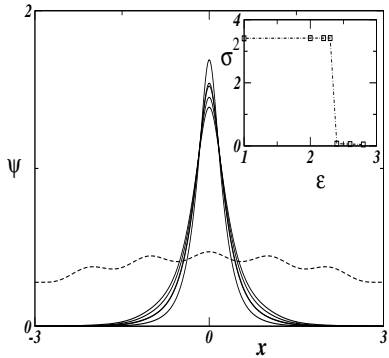}
\end{center}
\caption{The numerically found ground-state of the cubic-quintic GPE for
several values of $\protect\epsilon $. Solid lines, starting from the
narrowest wave function, refer to $\protect\epsilon =4.0,~3.0,~2.8,~2.6,~2.4$%
, and the dashed line refers to $\protect\epsilon =2.3$.  Parameters are $c=3.65$, $g=1$, $q=3$, $L=5$. 
Inset: the squared
width of the ground-state versus $\protect\epsilon $ (the dot-dashed line is
a guide to the eye). Critical value $\protect\epsilon _{\mathrm{crit}}$
obtained from the numerical data is $\protect\epsilon _{\mathrm{crit}%
}=2.35(5)$, which should be compared to the variational prediction given by
Eq. (\protect\ref{epsilon_critical}), which is $\protect\epsilon _{\mathrm{%
crit}}\simeq 2.88$.}
\label{deloc-2}
\end{figure}

\begin{figure}[t]
\begin{center}
\includegraphics[width=6.cm,height=6.cm,angle=270,clip]{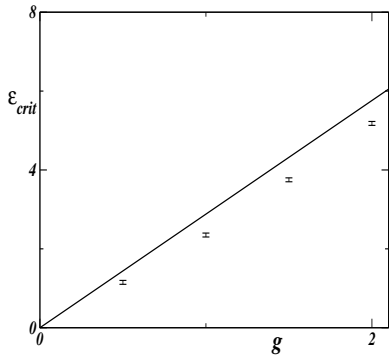}
\end{center}
\caption{Solid line: $\protect\epsilon _{\mathrm{crit}}$ versus $g,$ as
given by Eq. (\protect\ref{epsilon_critical}). Symbols refer to results
obtained from the numerical solution for the ground-state of the
cubic-quintic GPE. They represent the delocalization transition. The
parameters are the same as in Fig. (\protect\ref{deloc-2}).}
\label{comparison_epsilon}
\end{figure}

In Fig. \ref{deloc-2} we plot the numerically found ground-state of CQ GPE (%
\ref{cubic-quintic-mu}) for several values of $\epsilon $. It is seen that,
at small $\epsilon $, the wave function $\psi $ remains delocalized, until a
critical value is reached. In the inset of Fig. \ref{deloc-2} the squared
width of the numerically generated ground-state is plotted versus $\epsilon $%
. In Fig. \ref{comparison_epsilon}, we compare critical value $\epsilon _{%
\mathrm{crit}}$, as given by Eq. (\ref{epsilon_critical}), with numerical
results: for small $g$, the predicted linear dependence of $\epsilon _{%
\mathrm{crit}}$ on $g$ is well corroborated by the numerical results, the
relative error in the slope being $\sim 20\%$. In principle, the comparison
between variational estimate (\ref{epsilon_critical}) and numerical results
might be further improved by choosing a variational wave function which, in
the limit of $\epsilon =0$ (uniform space) would reproduce exact CQ soliton (%
\ref{cubic-quintic-1}). However, the calculations with such an ansatz are
extremely cumbersome.

\section{The effect of the harmonic trap}

\label{harmonic}

In this section we aim to use the variational approximation based on ansatz (%
\ref{VAR}) for examining the combined effect of the parabolic trapping
potential acting along with an OL, i.e., we take Eq. (\ref{quintic-mu}) with
external potential%
\begin{equation}
V_{\mathrm{ext}}(x)=\omega ^{2}x^{2}/2+\epsilon \sin ^{2}{(qx+\delta )},
\label{full-V}
\end{equation}%
cf. Eq. (\ref{OL}), and disregard binary collisions ($g=0$). Value $\delta =0
$ ($\delta =\pi $) corresponds to the matching (largest mismatch) between
the minimum of the harmonic potential and a local minimum of the lattice
potential. The respective variational energy is obtained from (\ref%
{E-functional}) with potential (\ref{full-V}):
\begin{equation}
E=\frac{\beta }{\sigma ^{2}}+\frac{\pi ^{2}\omega ^{2}\sigma ^{2}}{8}+\frac{%
\epsilon }{2}\left[ 1-\cos {(2\delta )}\mathrm{sech}(\pi q\sigma )\right] .
\label{E3-tot}
\end{equation}

With $\cos (2\delta )\geq 0$, the soliton is stable for $c<c^{\ast }$, and
it collapses otherwise. With $\cos (2\delta )<0$, a richer behavior is
predicted by the VA. The system does stabilize for $c<c^{\ast }$, while, for
$c>c^{\ast }$, the presence of the mismatched harmonic trap gives rise to a
metastability region. Since $E\rightarrow -\infty $ as $\sigma \rightarrow 0$
and $E\rightarrow +\infty $ as $\sigma \rightarrow \infty $, one can
encounter two possibilities: either $\partial E/\partial \sigma $ is
positive for all $\sigma $ (and there are no energy minima), or equation $%
\partial E/\partial \sigma =0$ has two roots, corresponding to a local
minimum and a maximum. The equation for the value of $\sigma $ at which
energy (\ref{E3-tot}) reaches the local minimum is
\begin{equation}
\left\vert \beta \right\vert =\frac{\epsilon \left\vert \cos \left( 2\delta
\right) \right\vert }{4\pi ^{2}q^{2}}\ell (\theta ),  \label{cond-trap}
\end{equation}%
where $\theta \equiv \pi q\sigma $, and
\begin{equation}
\ell (\theta )\equiv \theta ^{3}\left( \frac{\sinh {\theta }}{\cosh ^{2}{%
\theta }}-\eta \theta \right) ,  \label{ell}
\end{equation}%
\begin{equation}
\eta \equiv \frac{\omega ^{2}}{2\epsilon q^{2}\mid cos{(2\delta )}\mid }.
\label{eta}
\end{equation}%
One can see that, for $c=c^{\ast }$ (i.e., $\beta =0$), Eq. (\ref{cond-trap}%
) does not have a nonvanishing solution if $q$ is smaller than a critical
value,
\begin{equation}
q^{\left( \mathrm{cr}\right) }=\frac{\omega }{\sqrt{2\epsilon \left\vert
\cos \left( 2\delta \right) \right\vert }},
\end{equation}%
while it has a nonvanishing solution for $q>q^{\left( \mathrm{cr}\right) }$.

\begin{figure}[t]
\begin{center}
\includegraphics[width=6.cm,height=6.cm,angle=270,clip]{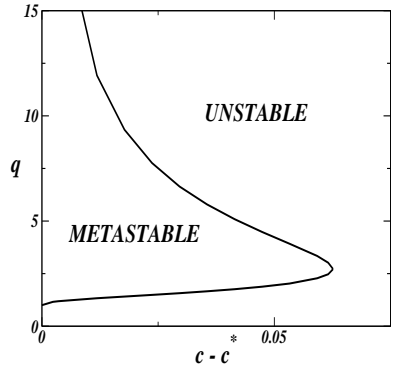}
\end{center}
\caption{The critical line separating in the $\left( q,c-c^{\ast }\right) $
plane of the model (including the parabolic trap) the metastable region from
the unstable one. The parameters are $\protect\epsilon =1$, $\cos {(2\protect%
\delta )}=-0.5$, and $\protect\omega =1$.}
\label{q}
\end{figure}

Actually, for $c>c^{\ast }$ (i.e., $\beta <0$), Eq. (\ref{cond-trap}) with $%
q>q^{\left( \mathrm{cr}\right) }$ has two nonvanishing roots, one of which
is a local minimum, while such roots do not exist for $q<q^{\left( \mathrm{cr%
}\right) }$. For $q>q^{\left( \mathrm{cr}\right) }$, the right-hand side of
Eq. (\ref{cond-trap}) has a maximum value, which fixes the maximum value of $%
\beta $, i.e., the maximum value of $c$, which we refer to as $c^{\ast \ast
\ast }$. Then, for $c>c^{\ast \ast \ast }$, the variational energy does not
have a local minimum. For $c^{\ast }<c<c^{\ast \ast \ast }$ there appears a
finite metastability region, in terms of wavenumber $q$, as illustrated by
Fig. \ref{q}. In other words, for fixed $c$, metastable states appear at
large values of $\epsilon $.

\section{Conclusions}

\label{conclusions}

In this work we have studied the effect of the OL (optical lattice) on the
1D Bose gas with attractive three-body and repulsive two-body interactions,
described by the GPE (Gross-Pitaevskii equation) with the CQ (cubic-quintic)
nonlinearity. Actually, the effective quintic attractive term in the GPE may
be induced by the residual
deviation of the condensate, tightly trapped in a cigar-shaped confining
potential, from the one-dimensionality (when the three body
losses are negligible) \cite{nearly1D,khaykovich06} 
or by three-body interaction terms between atoms according to 
recent proposals \cite{paredes07,buchler07}.

In the absence of an external potential, soliton solutions to this equation
with the CQ nonlinearity are known in the exact form, but they all are
strongly unstable. We have demonstrated that the OL opens a stability window
for the solitons, provided that the OL strength, $\epsilon $, exceeds a
finite minimum value. The size of the stability window depends on $\epsilon
/q^{2}$, where $q$ is the OL's wavenumber. We have also considered effects
of the additional harmonic trap, finding that, if the quintic nonlinearity
is strong enough ($c\geq c^{\ast }$), a metastability region may arise,
depending on the mismatch between minima of the periodic potential and
harmonic trap.

\appendix

\section{Localized solutions of the cubic-quintic Gross-Pitaevskii equation}

\begin{figure}[t]
\begin{center}
\includegraphics[width=6.cm,height=6.cm,angle=270,clip]{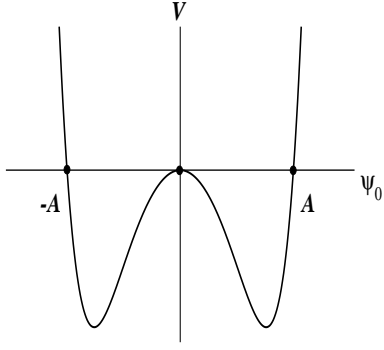}
\end{center}
\caption{Plot of potential $V(\protect\psi _{0})$ for $\protect\mu <0$.}
\label{potential}
\end{figure}

Assuming that $\psi (x)$ is real, we look for localized solutions to the CQ
NLS equation,
\begin{equation}
-\frac{1}{2}\frac{d^{2}\psi }{dx^{2}}+g\psi ^{3}-c\psi ^{5}=\mu \psi
\label{cubic-quintic-mu-app}
\end{equation}%
with $c>0$ and $g\geq 0$. Interpreting $x$ as a formal time variable and $%
\psi (x)$ as the coordinate of a particle, Eq. (\ref{cubic-quintic-mu-app})
formally corresponds to the Newton's equation of motion of this particle,
\begin{equation}
M\frac{d^{2}\psi }{dx^{2}}=-\frac{\partial V}{\partial \psi },
\end{equation}%
where the effective mass is $M=1/2$, and the potential is
\begin{equation}
V(\psi )=\frac{\mu }{2}\psi ^{2}-\frac{g}{4}\psi ^{4}+\frac{c}{6}\psi ^{6},
\label{potential-app}
\end{equation}%
with an arbitrary additive constant chosen so as to have $V(0)=0$. Potential
(\ref{potential-app}) for $\mu <0$, which corresponds to normalizable
solutions, is plotted in Fig. \ref{potential}. Condition $V(\pm A)=0$ yields
expression (\ref{A-cubic-quintic}) for the soliton's amplitude.

Further, we make use of the conservation of the corresponding Hamiltonian,
\begin{equation}
H=\frac{M}{2}\left( \frac{d\psi }{dx}\right) ^{2}+V(\psi ).
\label{energy-app}
\end{equation}%
The boundary conditions for localized solutions, $\psi (x\rightarrow \infty
)\rightarrow 0$, $d\psi /dx(x\rightarrow \infty )\rightarrow 0$, select $H=0$
in Eq. (\ref{energy-app}). Taking into regard the fact that $V(\psi )\leq 0$
for $0\leq \psi \leq A$, and looking for solutions with $d\psi /dx<0$ at $x>0
$, one obtains from here the soliton solution in an implicit
form,
\begin{equation}
x=\int_{\psi (x)}^{A}\frac{d\psi }{2\sqrt{-V(\psi )}}.  \label{quadr-app-1}
\end{equation}%
It further follows from Eq. (\ref{quadr-app-1}) that%
\begin{equation}
\mathcal{E}=\frac{\psi ^{2}(x)}{A^{2}}\frac{2a^{2}+b^{2}A^{2}}{%
2a^{2}+b^{2}\psi ^{2}(x)+2a\sqrt{a^{2}+b^{2}\psi ^{2}(x)-\psi ^{4}(x)}},
\label{quadr-app-2}
\end{equation}%
with $\mathcal{E}\equiv e^{-2\sqrt{2|\mu |}x}$. In Eq. (\ref{quadr-app-2}),
we use notation $a^{2}=3|\mu |/c$ and $b^{2}=3g/2c$. Thus, from Eq. (\ref%
{quadr-app-2}) one obtains
\begin{equation}
\psi ^{2}(x)=\frac{4a^{2}A^{2}\left( 2a^{2}+b^{2}A^{2}\right) \mathcal{E}}{%
\left[ 2a^{2}+b^{2}A^{2}\left( 1-\mathcal{E}\right) \right] ^{2}+4a^{2}A^{4}%
\mathcal{E}^{2}}.  \label{quadr-app-3}
\end{equation}%
One can easily check that this expression yields $\psi ^{2}(x)=A^{2}/\cosh {%
(2\sqrt{2|\mu |}x)}$ for $g=0$, and that $\psi (0)=A$, as it must be.
Finally, using relation $a^{2}+b^{2}A^{2}=A^{4}$, one obtains Eq. (\ref%
{cubic-quintic-1}) from Eq. (\ref{quadr-app-3}), after a straightforward
algebra.

\section{The variational energy}

\begin{figure}[t]
\begin{center}
\includegraphics[width=6.cm,height=6.cm,angle=270,clip]{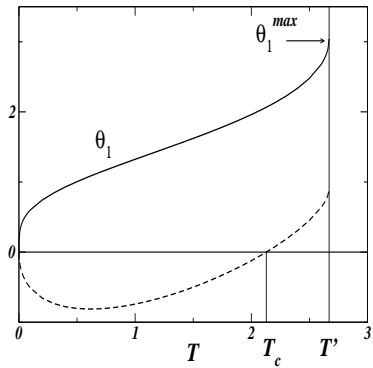}
\end{center}
\caption{The solid line represents 
$\protect\theta_{1}$ as a function of parameter $T$
(defined in Eq. (\protect\ref{T-cond})) for $g=0$ . The dashed line is 
the 
plot of function $T-2\protect\theta _{1}^{2}(T)/\cosh {\protect\theta _{1}(T)%
}$ versus $T$. The maximum value of $\protect\theta _{1}$ at $T=T^{\prime }$
is indicated.}
\label{fig_theta}
\end{figure}

In this Appendix we aim to study minima of variational energy (\ref{E3}).
When $g=0$, one sees that, for $c>c^{\ast }$, the energy per particle tends
to $-\infty $ at $\sigma \rightarrow 0$, and to $\epsilon /2$ at $\sigma
\rightarrow \infty $. Then, with regard to $\partial E/\partial \sigma >0$,
no local (metastable) minima exist, and variational wave function (\ref{VAR}%
) is not the ground-state for any finite width. For $c=c^{\ast }$, one
obtains the global minimum at $\sigma =0$, which implies the collapse. For $%
c<c^{\ast }$, the situation is different: $E\rightarrow \infty $ as $\sigma
\rightarrow 0$ (because $\beta >0$), and $E-\epsilon /2\rightarrow +0$ for $%
\sigma \rightarrow \infty $. Then, it is necessary to find the value of $%
\beta $ at which derivative $\partial E/\partial \sigma $ has two real
zeros. Introducing the parameter
\begin{equation}
T\equiv \frac{4\beta \pi ^{2}q^{2}}{\epsilon },  \label{T}
\end{equation}%
with $\beta $ defined as per Eq. (\ref{beta}), one can write condition $%
\partial E/\partial \sigma =0$ as
\begin{equation}
T=\theta ^{3}\frac{\sinh {\theta }}{\cosh ^{2}{\theta }},  \label{T-cond}
\end{equation}%
where $\theta =\pi q\sigma $, as defined above. Equation (\ref{T-cond}) can
be satisfied if $T$ is smaller than a maximum value, 
$T^{\prime }\approx 2.67$, 
and it then has two roots, $\theta _{1}$ and $\theta _{2}$, which
correspond, respectively to the minimum at $\sigma =\sigma _{1}$, and
maximum at $\sigma =\sigma _{2}$ (see Fig. \ref{fig1}). For $T>T^{\prime }$,
Eq. (\ref{T}) has no roots, hence the variational energy has no minima at
finite values of the soliton's width, $\sigma $. A plot of $\theta _{1}$ as
a function of $T$ is presented in Fig. \ref{fig_theta}, where the maximum
value of $\theta _{1}$ is $\theta _{1}^{\max }\approx 3.0415$. The energy
minimum at $\theta _{1}$ is a global one if $E(\theta _{1})<\epsilon /2$;
using Eq. (\ref{E3}), this condition reads
\begin{equation}
T-\frac{2\theta _{1}^{2}(T)}{\cosh {\theta _{1}(T)}}<0.
\label{condition-app}
\end{equation}%
As one can see from Fig. \ref{fig_theta}, condition (\ref{condition-app}) is
satisfied for $T<T_{c}$, where $T_{c}\approx 2.1289$; then, a global minimum
exists only for $0<T<T_{c}$, while for $T_{c}<T<T^{\prime }$ the minimum is
local, corresponding to a metastable state. Using the value of $T_{c}$ and
definition (\ref{T}), one arrives at Eq. (\ref{c_2_star}).

For $g>0$ (recall it corresponds to the two-body repulsion), variational
energy (\ref{E3}) for $c>c^{\ast }$ does not have a minimum at finite values
of $\sigma $. However, for $c=c^{\ast }$ a finite minimum is possible.
Indeed, with definition of $G$ as per Eq. (\ref{G-cond-text}), condition $%
\partial E/\partial \sigma =0$ can be written as
\begin{equation}
G=\theta ^{2}\frac{\sinh {\theta }}{\cosh ^{2}{\theta }}.  \label{G-cond}
\end{equation}%
For $G<G^{\prime }\approx 1.0341$, Eq. (\ref{G-cond}) has two roots. By
imposing the condition that the value of the energy at $\sigma =\sigma _{1}$
be smaller than $\epsilon /2$, one gets $G<G_{c}\simeq 0.6627$. Then,
similar to the situation considered above, a global minimum exists only for $%
0<G<G_{c}$, while for $G_{c}<G<G^{\prime }$ the minimum is local.

\begin{figure}[t]
\begin{center}
\includegraphics[width=6.cm,height=6.cm,angle=270,clip]{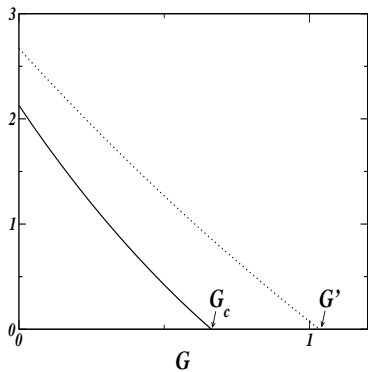}
\end{center}
\caption{The solid (dashed) line is the plot of $T_{c}$ ($T^{\prime }$) as a
function of parameter $G$.}
\label{T-vs-G}
\end{figure}

\begin{figure}[t]
\begin{center}
\includegraphics[width=6.cm,height=6.cm,angle=270,clip]{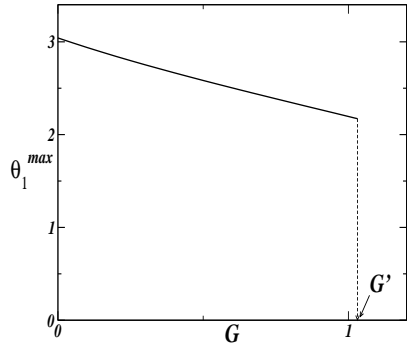}
\end{center}
\caption{Plot of maximum value $\protect\theta _{1}^{\max }$ of the
smaller root of Eq. (\protect\ref{T-cond}), $\protect\theta _{1}$ (at $%
T=T^{\prime }(G)$), as a function of $G$.}
\label{theta-max}
\end{figure}

For $c<c^{\ast }$, condition $\partial E/\partial \sigma =0$ reads
\begin{equation}
T=\theta ^{3}\frac{\sinh {\theta }}{\cosh ^{2}{\theta }}-G\theta .
\label{T-G-cond}
\end{equation}%
One can see that condition (\ref{T-G-cond}) is satisfied for $T<T^{\prime
}(G)$, with $T^{\prime }(G^{\prime })=0$. Then, for $G>G^{\prime }$, i.e.,
for $\epsilon $ small enough, the variational energy does not have a
minimum. Imposing the condition that the minimum is global leads to $T<T_{c}$%
, with $T_{c}(G_{c})=0$. Then, for $G>G_{c}$, i.e. for $\epsilon $ smaller
than a critical value, the variational energy cannot have a \emph{global}
minimum at a finite value of $\sigma $, i.e., localized states cannot
realize a global minimum. Functions $T^{\prime }(G)$ and $T_{c}(G)$ are
plotted in Fig. \ref{T-vs-G}; in Fig. \ref{theta-max}, we plot maximum value
$\theta _{1}^{\max }$ of $\theta _{1}$ for $T=T^{\prime }(G)$, as a function
of $G$.

\emph{Acknowledgement:} We thank L. Salasnich for useful comments
and discussions. This work was partially supported by ESF project
INSTANS and by MIUR projects ``Quantum Field Theory and
Statistical Mechanics in Low Dimensions", and ``Quantum Noise in
Mesoscopic Systems".



\begin{thebibliography}{99}
\bibitem{book1} C. J. Pethick and H. Smith, \textit{Bose-Einstein
Condensation in Dilute Alkali Gases} (Cambridge, Cambridge University Press,
2002)

\bibitem{book2} L. P. Pitaevskii and S. Stringari, \textit{Bose-Einstein
Condensation} (Oxford, Clarendon Press, 2003)

\bibitem{morsch06} O. Morsch and M. K. Oberthaler, Rev. Mod. Phys. \textbf{78%
}, (2006) 179

\bibitem{jaksch98} D. Jaksch, C. Bruder, J. I. Cirac, C. W. Gardiner, and P.
Zoller, Phys. Rev. Lett. \textbf{81}, (1998) 3108

\bibitem{hansch02} M. Greiner, O. Mandel, T. Esslinger, T. W. H\"ansch, and
I. Bloch, Nature \textbf{419}, (2002) 51

\bibitem{trombettoni01} A. Trombettoni and A. Smerzi, Phys. Rev. Lett.
\textbf{86}, (2001) 2353

\bibitem{abdullaev01} F. Kh. Abdullaev, B. B. Baizakov, S. A. Darmanyan, V.
V. Konotop, and M. Salerno, Phys. Rev. A \textbf{64}, (2001) 043606

\bibitem{alfimov02} G. L. Alfimov, P. G. Kevrekidis, V. V. Konotop, and M.
Salerno, Phys. Rev. E \textbf{66}, (2002) 046608

\bibitem{Mason} M. A. Porter, R. Carretero-Gonz\'{a}lez, P. G. Kevrekidis,
and B. A. Malomed, Chaos \textbf{15}, (2005) 015115.

\bibitem{wu01} B. Wu and Q. Niu, Phys. Rev. A \textbf{64}, (2001) 061603(R)

\bibitem{smerzi02} A. Smerzi, A. Trombettoni, P. G. Kevrekidis, and A. R.
Bishop, Phys. Rev. Lett. \textbf{89}, (2002) 170402

\bibitem{konotop02} V. V. Konotop and M. Salerno, Phys. Rev. A \textbf{65},
(2002) 021602

\bibitem{wu03} B. Wu and Q. Niu, New J. Phys. \textbf{5}, (2003) 104

\bibitem{menotti03} C. Menotti, A. Smerzi, and A. Trombettoni, New J. Phys.
\textbf{5}, (2003) 112

\bibitem{taylor03} E. Taylor and E. Zaremba, Phys. Rev. A \textbf{68},
(2003) 053611

\bibitem{kramer03} M. Kr\"amer, C. Menotti, L. P. Pitaevskii, and S.
Stringari, Eur. Phys. J. D \textbf{27}, (2003) 247

\bibitem{kramer05} M. Kr\"amer, C. Menotti and M. Modugno, J. Low Temp.
Phys. \textbf{138}, (2005) 729

\bibitem{cataliotti03} F. S. Cataliotti, L. Fallani, F. Ferlaino, C. Fort,
P. Maddaloni, and M. Inguscio, New J. Phys. \textbf{5}, (2003) 71

\bibitem{inguscio04} L. Fallani, L. De Sarlo, J. E. Lye, M. Modugno, R.
Saers, C. Fort, and M. Inguscio, Phys. Rev. Lett. \textbf{93}, (2004) 140406

\bibitem{GSprediction} F. Kh. Abdullaev, B. B. Baizakov, S. A. Darmanyan, V.
V. Konotop, and M. Salerno, Phys. Rev. A \textbf{64}, 043606 (2001); I.
Carusotto, D. Embriaco, and G. C. La Rocca, \textit{ibid}. \textbf{65},
(2002) 053611; B. B. Baizakov, V. V. Konotop, and M. Salerno, J. Phys. B
\textbf{35}, (2002) 5105; E. A. Ostrovskaya and Y. S. Kivshar, Phys. Rev.
Lett. \textbf{90}, (2003) 160407; Opt. Exp. \textbf{12}, (2004) 19 

\bibitem{eiermann04} B. Eiermann, T. Anker, M. Albiez, M. Taglieber, P.
Treutlein, K. P. Marzlin, and M. K. Oberthaler, Phys. Rev. Lett. \textbf{92}%
, (2004) 230401

\bibitem{strecker02} K. E. Strecker, G. B. Partridge, A. G. Truscott, and R.
G. Hulet, Nature \textbf{417}, (2002) 150

\bibitem{khaykovich02} L. Khaykovich, F. Schreck, G. Ferrari, T. Bourdel, J.
Cubizolles, L. D. Carr, Y. Castin, and C. Salomon, Science \textbf{296},
(2002) 1290

\bibitem{Weiman} S. L. Cornish, S. T. Thompson, and C. E. Wieman, Phys. Rev.
Lett. \textbf{96}, (2006) 170401

\bibitem{malomed06} B. A. Malomed, \textit{Soliton Management in Periodic
Systems} (Springer-Verlag, New York, 2006)

\bibitem{sulem99} C. Sulem and P.-L. Sulem, \textit{The Nonlinear Schr\"{o}%
dinger Equation} (Springer-Verlag, New York, 1999)

\bibitem{ablowitz04} M. J. Ablowitz, B. Prinari, and A. D. Trubatch, \textit{%
Discrete and Continuous Nonlinear Schr\"{o}dinger Systems} (Cambridge,
University Press, 2004)

\bibitem{malomed03} B. B. Baizakov, B. A. Malomed, and M. Salerno, Europhys.
Lett. \textbf{63}, (2003) 642

\bibitem{Yang} J. Yang and Z. H. Musslimani, Opt. Lett. \textbf{28}, (2003)
2094

\bibitem{BBB} B. B. Baizakov, B. A. Malomed, and M. Salerno, Phys. Rev. A
\textbf{70}, (2004) 053613; Phys. Rev. E \textbf{74}, (2006) 066615

\bibitem{Barcelona} D. Mihalache, D. Mazilu, F. Lederer, Y. V. Kartashov,
L.-C. Crasovan, and L. Torner, Phys. Rev. E \textbf{70}, (2004) 055603(R)

\bibitem{paredes07} B. Paredes, T. Keilmann, and J. I. Cirac, Phys. Rev. A
\textbf{75}, (2007) 053611

\bibitem{buchler07} H. P. Buchler, A. Micheli, and P. Zoller, Nature Phys.
\textbf{3}, (2007) 726

\bibitem{moore91} G. Moore and N. Read, Nucl. Phys. B \textbf{360}, (1991)
362

\bibitem{fersino08} E. Fersino, G. Mussardo, and A. Trombettoni, Phys. Rev.
A \textbf{77}, (2008) 053608

\bibitem{abdullaev05} F. Kh. Abdullaev and M. Salerno, Phys. Rev. A 
\textbf{72}, (2005) 033617

\bibitem{nearly1D} A. E. Muryshev, G. V. Shlyapnikov, W. Ertmer, K.
Sengstock, and M. Lewenstein, Phys. Rev. Lett. \textbf{89}, (2002) 110401

\bibitem{khaykovich06} L. Khaykovich and B. A. Malomed, Phys. Rev. A \textbf{%
74}, (2006) 023607

\bibitem{kivshar03} Yu. S. Kivshar and G. P. Agrawal, \textit{Optical
Solitons} (Elsevier Science, San Diego, 2003).

\bibitem{Radik2} R. Driben and B. A. Malomed, Eur. Phys. J. D, in press
(2008) (DOI: 10.1140/epjd/e2008-00239-3).

\bibitem{Radik1} R. Driben, B. A. Malomed, A. Gubeskys, and J. Zyss, Phys.
Rev. E \textbf{76}, (2007) 066604

\bibitem{malomed99} B. A. Malomed, Z. H. Wang, P. L. Chu, and G. D. Peng, J.
Opt. Soc. Am. B \textbf{16}, (1999) 1197

\bibitem{xu05} Z. Xu, Y. V. Kartashov, and L. Torner, Phys. Rev. Lett.
\textbf{95}, (2005) 113901

\bibitem{wang08} Z. Dai, Y. Wang, and Q. Guo, Phys. Rev. A \textbf{77},
(2008) 063834

\bibitem{malomed01} B. A. Malomed, Progr. Opt. \textbf{43}, (2001) 69

\bibitem{lieb63} E. H. Lieb and W. Liniger, Phys. Rev. \textbf{130}, (1963)
1605

\bibitem{mcguire64} J. B. McGuire, J. Math. Phys. \textbf{5}, (1964) 622

\bibitem{calogero75} F. Calogero and A. Degasperis, 
Phys. Rev. A \textbf{11}, (1975) 265

\bibitem{ablowitz81} M. J. Ablowitz and H. Segur, \textit{Solitons and the
Inverse Scattering Transform} (SIAM, Philadelphia, 1981)

\bibitem{polyanin04} A. D. Polyanin and V. F. Zaitsev, \textit{Handbook of
Nonlinear Partial Differential Equations} (Chapman \& Hall/CRC Press, Boca
Raton, 2004)

\bibitem{gaididei99} Yu. B. Gaididei, J. Schjodt-Eriksen, and P. L.
Christiansen, Phys. Rev. E \textbf{60}, (1999) 4877

\bibitem{pushkarov79} Kh. I. Pushkarov, D. I. Pushkarov, and I. V. Tomov,
Opt. Quantum Electron. \textbf{11}, (1979) 471

\bibitem{cowan86} S. Cowan, R. H. Enns, S. S. Rangnekar, and S. S. Sanghera,
Can. J. Phys. \textbf{64}, (1986) 311

\bibitem{kolokolov73} M. G. Vakhitov and A. A. Kolokolov, Radiophys. Quantum
Electron. \textbf{16}, (1973) 783

\bibitem{baym96} G. Baym and C. J. Pethick, Phys. Rev. Lett. \textbf{76},
(1996) 6

\bibitem{fetter95} A. L. Fetter, arXiv:cond-mat/9510037.

\bibitem{ruprecht95} P. A. Ruprecht, M. J. Holland, K. Burnett, and M.
Edwards, Phys. Rev. A \textbf{51}, (1995) 4704

\bibitem{Anderson} D. Anderson, Phys. Rev. A \textbf{27}, (1983) 3135

\bibitem{delocalization} B. B. Baizakov and M. Salerno, Phys. Rev. A \textbf{%
69}, (2004) 013602

%
%
\end{thebibliography}
\end{document}